\documentclass[aps,preprint]{revtex4-1}
\usepackage{graphicx,psfrag}
\usepackage{graphicx} 
\usepackage{bm}
\usepackage{booktabs}
\usepackage{amsmath,amsthm,amssymb}
\usepackage{gt12}
\usepackage{epstopdf}

\usepackage{color}
\begin{document}

\newcommand{\vso}{v_{\rm so}}
\newcommand{\EvR}{\Ev_{\rm R}}
\newcommand{\rhom}{\rho_{\rm m}}
\newcommand{\jvm}{\jv_{\rm m}}

\title{
Spin motive force induced by Rashba interaction in the strong sd coupling regime
      }
\author{Gen Tatara$^1$,$^2$, Noriyuki Nakabayashi$^1$, and Kyun-Jin Lee$^3$,$^4$
}  
\affiliation{ $^1$ Graduate School of Science and Engineering, Tokyo Metropolitan University, Hachioji, Tokyo 192-0397 Japan
 \\ $^2$ RIKEN Advanced Science Institute, 
2-1 Hirosawa, Wako, Saitama 351-0198, Japan
\\$^3$ Department of Materials Science and Engineering, Korea University, Seoul 136-701, Korea
\\$^4$ Spin Convergence Research Center, Korea Institute of Science and Technology, Seoul 136-791,
Korea
} 
\date{\today}
\begin {abstract} 
Spin motive force induced by the Rashba interaction in the presence of strong sd interaction between conduction electron and localized spin is theoretically studied.  
The motive force is calculated by evaluating the time-derivative of the current density on the basis of microscopic formalism.
It is shown that there are two motive forces, one proportional to $\EvR\times\dot{\nv}$, the other, perpendicular component, proportional to 
$\EvR\times(\nv\times\dot{\nv})$, where $\EvR$ and $\nv$ are the Rashba electric field and localized spin direction, respectively.
The second type arises in the strong sd coupling regime from the spin relaxation. 
The appearance of perpendicular component from the spin relaxation is understood from the analogy with the current-induced torques.
In the case of domain wall motion, the two contributions to the spin motive force are the same order of magnitude, while the first term dominates in the case of precession of uniform magnetization.
Our result explains appearance of the perpendicular component in the weak sd coupling limit, recently discussed in the context of spin damping monopole.
Detection of AC voltage induced by the precession of uniform magnetization serves as a experimental evidence of the Rashba interaction in films and wires. 

\end{abstract}

\maketitle

\section{Introduction}

In electronics, electric motive force driving electron's charge is used.
Recently, control of electron's spin has been carried out in spintronics technology, and use of another motive force, spin motive force, has become possible.
The idea of spin motive force induced by an sd interaction between conduction electron and localized spin was argued by Berger in the case of a domain wall motion in 1986 \cite{Berger86}.
In the strong sd coupling regime, the spin motive force is described by a U(1) gauge theory for an adiabatic component, $\As{\mu}{z}$, 
of the spin gauge field, $\As{\mu}{\alpha}$ ($\mu=x,y,z,t$ and $\alpha=x,y,z$ are indices for space time and spin) \cite{Volovik87}.
In the adiabatic limit and in the absence of spin relaxation, 
$\As{\mu}{z}$ has a U(1) gauge symmetry, and the spin electric field, defined by spin motive force divided by the electron charge ($e$),  is 
$\Es{i}=\partial_t \As{i}{z}-\partial_i \As{0}{z}$ \cite{Volovik87,Barnes07}.
By use of a unit vector, $\nv(\rv,t)$, describing the local spin direction, it is 
$\Es{i}=-\frac{\hbar}{2e}\nv\cdot(\dot{\nv}\times\partial_i\nv)$.
This spin electric field was experimentally observed in the case of moving domain wall \cite{Yang09,Hayashi12}. 

When there is spin relaxation, the spin motive force is modified.
In the context of current-driven torque, spin relaxation induces a perpendicular component, $\beta_{\rm sr}(\nv\times(\jv\cdot\nabla)\nv)$, to the adiabatic spin-transfer torque $(\jv\cdot\nabla)\nv$ ($\jv$ is the applied current density and $\beta_{\rm sr}$ is a coefficient proportional to the spin relaxation rate) \cite{Zhang04,Thiaville05}.
As discussed by Duine \cite{Duine09} and Shibata \cite{Shibata09}, spin motive force is the inverse effect of current-induced torque.
The spin relaxation is thus expected to modify spin electric field to give rise to a correction, 
 $E_i^{\rm (sr)} \propto \beta_{\rm sr}\nv\cdot[\dot{\nv}\times(\nv\times \nabla_i \nv)]
= \dot{\nv}\cdot\nabla_i\nv$ \cite{Tserkovnyak08,Duine09,Lucassen11}.

The above results apply irrespective of the origin of the spin relaxation; 
Relaxation due to spin flip scattering by magnetic defects and that due to spin-orbit interaction at heavy impurity sites result in the essentially the same result \cite{KTS06,TE08}.
In contrast, Rashba spin-orbit interaction, which arises from a breaking of the inversion symmetry, leads to a qualitatively different current-induced torque, and are predicted to realize very fast domain wall motion and to give rise to a spin Hall effect \cite{Obata08,Manchon09,Bijl12}. 
The effect of Rashba interaction on the spin electric field was studied very recently \cite{Takeuchi12,Kim12,GT12}.
Takeuchi et al. investigated weak sd coupling regime and found a spin electric field of 
${\Evs}=a[\EvR\times(\nv\times\dot{\nv})]\equiv \Ev_{s\perp}$, where $a$ is a constant and $\EvR$ is the electric field of Rashba interaction \cite{Takeuchi12}.
They also calculated the spin magnetic field, $\Bsv$, and found that the spin electromagnetic field has a monopole.
Namely, the field has a finite monopole density and monopole current density, defined in electromagnetism theory  as
$\rhom\equiv \nabla\cdot\Bvs$ and $\jvm\equiv \nabla\times {\Evs}+\dot{\Bvs}$, respectively.
The method employed by Takeuchi to address the spin electromagnetic field is unique; the fields were identified by calculating the induced electric current density and then comparing the result with a general expression, $\jv=\sigma {\Evs}+\nabla\times \Bvs -D\nabla\nel$, where $\sigma$ is conductivity, $D$ is diffusion constant and $\nel$ is charge density.
This approach is highly useful to study the weak coupling regime, where adiabatic component of spin gauge field can not be defined.
Kim et al. studied, on the other hand, a strong sd coupling regime and obtained a different form of 
${\Evs}=b[\EvR\times\dot{\nv}]
$ ($b$ is a constant) \cite{Kim12}.
Therefore, the spin electric fields so far identified are different for strong and weak coupling regimes.
The absence of $\Ev_{s\perp}$ in the strong coupling limit would be due to the fact that the effect of spin relaxation is not taken account of in Ref. \cite{Kim12}, as is suggested from the case of current-induced torques.

The aim of this paper is to study the spin electric field (motive force) induced by Rashba interaction and sd interaction on a basis of a microscopic theory  taking account of the spin relaxation due to spin-orbit interaction driven by heavy impurities.
Spin relaxation effect is expected to be essential in grasping global behavior of the spin electric field.
We have mentioned above two methods used for far to calculate effective spin electromagnetic field;
1) by use of adiabatic gauge field, and 2) by calculating electric current. 
Here we will use different approach. 
The motive force, $\Fv$, acting on conduction electrons is defined by an average of the time derivative of electron velocity multiplied by mass, $m$.
It is therefore calculated by evaluating $\del{\jv}{t}$ as
\begin{align}
\Fv=\frac{m}{e\nel}\del{\jv}{t},\label{Fdef}
\end{align}
 where $\nel$ is the electron density and $e$ is the electron charge.
This argument was employed in Ref. \cite{Kim12} for a phenomenological derivation of spin motive force.
We will estimate \Eqref{Fdef} microscopically by rewriting the time-derivative by a commutator of the current density operator, $\hat{\jv}$, and the Hamiltonian $H$ as 
\begin{align}
\Fv=\frac{im}{e\hbar\nel}\average{[H,\hat{\jv}]},
\label{Fdefcom}
\end{align}
where $\average{\ }$ denotes quantum average.
We will show that when spin relaxation is taken into account, $\Ev_{{\rm s},\perp}$ indeed emerges even in the strong sd coupling regime, as is consistent with previous analyses on current-induced torques \cite{Zhang04,Thiaville05,KTS06,TE08} and on spin motive force without Rashba interaction \cite{Duine09}.
In fact,  the Rashba interaction gives rise to a spin electric field 
\begin{align}
{\Evs}^{\rm R}=\gamma_{\rm R} \lt[ (\alphaRv \times\dot{\nv})
     + \beta_{\rm R} (\alphaRv \times(\nv\times\dot{\nv}) ) \rt] , \label{Evsintro}
\end{align}
where $\gamma_{\rm R}=\frac{m}{\hbar e}$ and  $\beta_{\rm R}$ represents the strength of the spin relaxation.
The first term was pointed out in Ref. \cite{Kim12}, and the second term has the form found in Ref. \cite{Takeuchi12} in the weak sd coupling limit.
In the case of precession of uniform magnetization, the first contribution dominates, since $\beta_{\rm R}$ is small (of the order of $0.01$), as discussed in Sec. \ref{SEC:dw}. 
For a moving domain wall in a perpendicular anisotropy system, the contributions from the first and second contributions of \Eqref{Evsintro} are the same order. For current-induced magnetization switching in a perpendicular magnetic random access memory, the first contribution enhances the damping and consequently, increases the threshold switching current density.

Based on a series of experiments on Pt/Co/AlO$_x$ systems \cite{Miron10,Miron11},  
Miron et al. argued that an injection of an in-plane current into
this system generates a large transverse magnetic field, predicted from
Rashba theories \cite{Obata08,Manchon09}. On the other hand, Liu et
al. \cite{Liu11,Liu12}
 reported that spin transfer
torques induced by an in-plane current in similar systems can be
quantitatively explained by the spin Hall effect and they could not find any
signatures of Rashba effect in their measurements. These contradictory
experiments have led to extensive discussions on the existence of Rashba
effect in ferromagnets. 

As we will show in Sec. \ref{SEC:dw},
the amplitude of Rashba-induced alternating spin voltage in the case of a precession of a single domain magnetization of a size of 100nm is predicted to be as large as $0.2$mV 
at a frequency of 100MHz if the Rashba coupling constant is 
$\alphaR=3\times 10^{-10}$eV$\cdot$m \cite{Kim12}.
From the Rashba spin motive force, \Eqref{Evsintro}, the Rashba coupling constant, $\alphaR$, is determined since the prefactor,  $\gamma_{\rm R}$, is universal and does not depends on material properties ($\beta_R$ is small and is neglected  in a uniform magnetization case). 
At the same time, $\alphaR$ is measured from a domain wall speed \cite{Miron10}. 
Measurement of AC spin voltage without domain walls is thus expected to provide a highly useful information concerning the existence of Rashba effect.

For confirmation of Rashba-induced spin motive force, it is essential to exclude similar signal from the spin pumping and inverse spin Hall effects \cite{Saitoh06}. 
In the case of perpendicular easy axis systems like in Ref. \cite{Miron10}, the dominating part of inverse spin Hall signal vanishes, since the polarization of spin current is parallel to the spin current flow.

In the next two sections, we show derivation of the Rashba-induced spin motive force.
Those who are interested in the result and its physical consequences only may skip the section and read Sec. \ref{SEC:dw}.

\section{Spin motive force without spin relaxation}

We first study the case without spin relaxation, studied in Ref. \cite{Kim12} by evaluating \Eqref{Fdefcom}.
The interactions we consider is the sd interaction and the Rashba interaction.
The Hamiltonian is
\begin{align}
H= \int d^3r c^\dagger \lt[
-\frac{\hbar^2\nabla^2}{2m}-\ef-\Deltasd \nv\cdot \sigmav +i\alphaRv\cdot(\nabla\times\sigmav) 
                   \rt] c,
\end{align}
where $\Deltasd$ is the energy splitting of the conduction electron, $\nv$ represents the direction of localized spin, $\alphaRv$ is a vector representing direction and strength of the Rashba interaction.
The electron is described by creation and annihilation operators, $\cdag$ and $c$, respectively.
These operators have two spin components as $c=(c_+,c_-)$ where $\pm$ represents the spin index. 
The electric current in the present system reads 
\begin{align}
\jv=-\frac{ie\hbar}{2m}\cdag \vvec{\nabla}c+ \frac{e}{\hbar}\cdag (\alphaRv\times\sigmav )c.
\end{align}
We consider a strong sd coupling limit, and carry out a local gauge transformation defining new electron operator by $a=Uc$, where $U$ is a $2\times2$ matrix. 
The matrix is chosen to diagonalized the sd interaction, i.e., 
$U=\mv\cdot\sigmav$, where $\mv=(\sin\frac{\theta}{2}\cos\phi,\sin\frac{\theta}{2}\sin\phi,\cos\frac{\theta}{2})$ ($\theta$ and $\phi$ are the polar coordinates of $\nv$) \cite{TKS_PR08}.
After the gauge transformation, the Hamiltonian and the current read
\begin{align}
H &= \int d^3r \lt[  a^\dagger \lt(-\frac{\hbar^2\nabla^2}{2m}-\ef-\Deltasd \sigmaz \rt) a  \rt. \nnr
& +\frac{i}{2}\lt(-\frac{\hbar^2\As{j}{l}}{m}
+\epsilon_{ijk}\alphaR _i R_{kl}\rt) \adag \vvec{\nabla}_j\sigma_l a
\nnr
& \lt. + \hbar \As{0}{l}\adag\sigma_l a
+\frac{1}{2}\lt(\frac{\hbar^2(A_{\rm s})^2}{m}-\epsilon_{ijk}\alphaR_i R_{kl}\As{j}{l}\rt) \adag a \rt],
\end{align}
where $\As{\mu}{l}\equiv\frac{1}{2}\tr[\sigma_l U^\dagger\partial_\mu U]=(\mv\times\partial_\mu\mv)_l$ 
is the spin gauge field and
$R_{kl}=2m_k m_l-\delta_{kl}$ is rotation matrix element.

By calculating the commutator $[H,\hat{\jv}]$ in \Eqref{Fdefcom}, we obtain the force operator as 
\begin{align}
\hat{\Fv} & = -\frac{\Deltasd}{\nel}\cdag \nabla(\nv\cdot\sigmav) c 
  +\frac{2m\Deltasd}{\hbar^2 \nel}\cdag [\alphaRv\times (\nv\times\sigmav) ] c+O((\alphaR)^2).
\label{forceoperator}
\end{align}
The first term containing $\nabla \nv$ is the term discussed in Refs. \cite{Berger78,TK04} in the context of current-driven domain wall motion.
Carrying out the gauge transformation and taking the field expectation value, the force reads
\begin{align}
F_i  = -\frac{\Deltasd}{\nel}
 \sum_{jn} \lt( (\nabla_i n_j) R_{jn}-\frac{2m}{\hbar^2} \sum_{klm} \epsilon_{ijk} \epsilon_{klm}\alphaR_j n_l R_{mn}\rt) 
\average{ \adag \sigma_n a}
+O(\alphaR^2). \label{Faftergaugetr}
\end{align}

\begin{figure}\centering
\includegraphics[width=0.25\hsize]{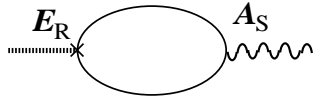}
\caption{
The Feynman diagrams describing the spin motive force force arising from the Rashba interaction at the lowest order.
The vertex with $\EvR$ represents the force induced by the Rashba field (the second term of \Eqref{forceoperator}), and the interaction with the spin gauge field, $\Av_{\rm s}$, is included to the linear order.
\label{FIGforce_A}}
\end{figure}
The contribution from the first term of \Eqref{Faftergaugetr} turns out to be the second order in $A_{\rm s}$, and is neglected in the adiabatic limit. 
The contribution from the second term, described in Fig. \ref{FIGforce_A},  is calculated as
\begin{align}
F_i  =  -i \frac{2m\Deltasd}{ \nel}
 \sum_{jn} \sum_{klm} \epsilon_{ijk} \epsilon_{klm} \alphaR_j n_l R_{mn} \sum_\alpha \As{0}{\alpha}
\sum_{\qv\Omega}e^{-i\qv\cdot\rv}e^{i\Omega t} \sum_{\kv\omega}
\tr[\sigma_n \gf{\kv\omega}{} \sigma_\alpha \gf{\kv+\qv,\omega+\Omega}{} ]^<,
\end{align}
where $\gf{\kv\omega}{}$ is the electron Green's function on the Keldysh contour with wave vector $\kv$ and angular frequency $\omega$, $^<$ denotes lesser component and $\sum_\omega\equiv\int\frac{d\omega}{2\pi}$.
We have neglected the contribution containing the spatial component of the spin gauge field, since they turn out to be order of $(A_{\rm s})^2$.
The lesser component is calculated as 
\begin{align}
\tr[\sigma_n \gf{\kv\omega}{ } \sigma_\alpha \gf{\kv+\qv,\omega+\Omega}{ } ]^<
& = -f_{\omega+\Omega} \tr[\sigma_n \gf{\kv\omega}{\ret} \sigma_\alpha \gf{\kv+\qv,\omega+\Omega}{\ret}]  \nnr
& +f_{\omega} \tr[\sigma_n \gf{\kv\omega}{\adv}\sigma_\alpha\gf{\kv+\qv,\omega+\Omega}{\adv}]  \nnr
& +(f_{\omega+\Omega}-f_{\omega}) \tr[\sigma_n \gf{\kv\omega}{\ret} \sigma_\alpha \gf{\kv+\qv,\omega+\Omega}{\adv}] .
\end{align}
Trace in the spin space is taken by use of 
\begin{align}
\tr[\sigma_n A \sigma_\alpha B]
=(\delta_{n\alpha}-\delta_{nz}\delta_{\alphaz})\sum_\pm A_\pm B_\mp
+ \delta_{nz}\delta_{\alphaz} \sum_\pm A_\pm B_\pm
-i\epsilon_{n\alpha z}\sum_\pm (\pm)A_\pm B_\mp,
\end{align}
where $A\equiv \lt(\begin{array}{cc} A_+ & 0 \\ 0 & A_- \end{array}\rt)$ and 
$B\equiv \lt(\begin{array}{cc} B_+ & 0 \\ 0 & B_- \end{array}\rt)$ are diagonal matrices.
The force is therefore given by
\begin{align}
F_i  =  \frac{2m\Deltasd}{\hbar\nel}
 \sum_{jn} \sum_{klm} \epsilon_{ijk} \epsilon_{klm} \alphaR_j n_l
\lt(  (R_{mn}\As{0}{n} -R_{mz}\As{0}{z}) \gamma_1^0 + R_{mz}\As{0}{z} \gamma_2^0 
-i\ \sum_\alpha \epsilon_{n\alpha z}R_{nm}\As{0}{\alpha} \gamma_3^0 \rt),
\end{align}
where
\begin{align}
\gamma_1^0 & \equiv -i\hbar \sum_{\kv\omega\pm} \lt[ 
 -f_{\omega+\Omega} \gf{\kv\omega\pm}{\ret}\gf{\kv+\qv,\omega+\Omega,\mp}{\ret}  
+f_{\omega} \gf{\kv\omega\pm}{\adv}\gf{\kv+\qv,\omega+\Omega,\mp}{\adv}     
 +(f_{\omega+\Omega}-f_{\omega}) \gf{\kv\omega\pm}{\ret}\gf{\kv+\qv,\omega+\Omega,\mp}{\adv}\rt]  
\nnr
\gamma_2^0 & \equiv -i\hbar\sum_{\kv\omega\pm} \lt[ 
 -f_{\omega+\Omega} \gf{\kv\omega\pm}{\ret}\gf{\kv+\qv,\omega+\Omega,\pm}{\ret} 
 +f_{\omega} \gf{\kv\omega\pm}{\adv}\gf{\kv+\qv,\omega+\Omega,\pm}{\adv}     +(f_{\omega+\Omega}-f_{\omega}) \gf{\kv\omega\pm}{\ret}\gf{\kv+\qv,\omega+\Omega,\pm}{\adv}\rt]  
\nnr
\gamma_3^0 & \equiv -i\hbar \sum_{\kv\omega\pm} (\pm) \lt[ 
 -f_{\omega+\Omega} \gf{\kv\omega\pm}{\ret}\gf{\kv+\qv,\omega+\Omega,\mp}{\ret} 
+f_{\omega} \gf{\kv\omega\pm}{\adv}\gf{\kv+\qv,\omega+\Omega,\mp}{\adv}   
 +(f_{\omega+\Omega}-f_{\omega}) \gf{\kv\omega\pm}{\ret}\gf{\kv+\qv,\omega+\Omega,\mp}{\adv}\rt]  .
\end{align}

The terms proportional to $\gamma_1^0$ and $\gamma_3^0$ are calculated using \cite{TKS_PR08}
\begin{align}
\sum_{lmn}  \epsilon_{klm}  n_l  (R_{mn}\As{0}{n} -R_{mz}\As{0}{z})
&= \frac{1}{2} \delp{n_k}{t},
\end{align}
and 
\begin{align}
\sum_{n\alpha}  \epsilon_{n\alpha z}  R_{mn}\As{0}{\alpha}
&= - \frac{1}{2} \delp{n_m}{t},
\end{align}
and the term proportional to $\gamma_2^0$ vanishes.
The result of the force is thus
\begin{align}
\Fv =  \frac{m\Deltasd}{\hbar \nel}
 \lt[     \gamma_1^0(\alphaRv \times \dot{\nv})
        +\gamma_3^0 (\alphaRv \times (\nv\times\dot{\nv})) \rt]
\end{align}

We evaluate the coefficient $\gamma_1^0$ and $\gamma_3^0$ in the clean limit (infinite elastic lifetime) 
and in the adiabatic limit (approximating $q=0$ in the Green's functions) using the identity
\begin{align}
\gf{\kv\omega\pm}{\adv}-\gf{\kv\omega\pm}{\ret}=2\pi i \delta(\hbar\omega-\epsilon_{\kv\pm}).
\end{align}
The result is
\begin{align}
\gamma_1^0 
&\simeq -\frac{1}{2\Deltasd} \sum_{\kv\pm}(\pm) \lt( f(\epsilon_{\kv\pm})-f(\epsilon_{\kv\mp}) \rt)
=-\frac{1}{\Deltasd}(n_+-n_-),
\end{align}
where $n_\pm=\sum_{\kv}f(\epsilon_{\kv\pm})$ is the density of spin $\pm$ electron,
and
\begin{align}
\gamma_3^0 
&\simeq -\frac{1}{2\Deltasd} \sum_{\kv\pm}\lt( f(\epsilon_{\kv\pm})-f(\epsilon_{\kv\mp}) \rt)
=0.
\end{align}

The motive force on the electron spin in the absence of spin relaxation is therefore obtained as
\begin{align}
\Fv =  -\frac{ms}{\hbar \nel}(\alphaRv \times \dot{\nv}), \label{Fbare}
\end{align}
where $s\equiv n_+-n_-$ is the spin polarization of the conduction electron. 
(In the strong coupling limit, $s=\nel$.)
The perpendicular component ($\alphaRv \times (\nv\times \dot{\nv})$) thus does not exist in the absence of spin relaxation, in agreement with the result in Ref. \cite{Kim12}.

\section{Effect of spin relaxation}

Here we study the effect of spin relaxation on the spin motive force.
The spin relaxation we consider is the one due to spin-orbit interaction due to heavy impurities, described by the following Hamiltonian: 
\begin{align}
\Hso= -\frac{i}{2} \int d^3r \sum_{ijk} \epsilon_{ijk} (\nabla_i \vso^{(k)}) 
    c^\dagger \vvec{\nabla}_j \sigma_k c,
\end{align}
where $\vso^{(k)}$ is random potential due to random impurities.
To obtain physical force, the potential is averaged over the random distribution, and the averaging is carried out taking account of the spin direction, denoted by the label $k$ as carried out in Ref. \cite{TE08}.
After the gauge transformation, the spin-orbit interaction reads
\begin{align}
\Hso= \int d^3r \sum_{ijk}  \epsilon_{ijk}  (\nabla_i \vso^{(k)}) R_{kl}
    \lt(-\frac{i}{2} a^\dagger \vvec{\nabla}_j \sigma_l a + \As{j}{l} a^\dagger a \rt),
\end{align}
and, in the Fourier representation, 
\begin{align}
\Hso & = -i \sum_{\kv\kv'\qv} \sum_{ijk}  \epsilon_{ijk}  \vso^{(k)}(\kv'-\kv-\qv)(k'-k-q)_i   \sum_{\omega\Omega}
\lt[ \frac{1}{2}(k'+k)_j R_{kl}(\qv,\Omega) a^\dagger_{\kv,\omega} \sigma_l a_{\kv',\omega+\Omega} 
  \rt. \nnr 
& \lt. 
    + \sum_{\qv'\Omega'} R_{kl}(\qv-\qv',\Omega-\Omega') \As{j}{l}(\qv',\Omega') 
           a^\dagger_{\kv,\omega} a_{\kv',\omega+\Omega}  
    \rt]. \label{HsoFT}
\end{align}
Here Fourier components are defined as 
$
\As{j}{l}(\qv,\Omega) = \int d^3r \int dt e^{i\qv\cdot\rv}e^{-i\Omega t} \As{j}{l} 
$ and 
$\vso(\qv) = \int d^3r  e^{i\qv\cdot\rv} \vso(\rv)$.
We assume that the impurities are heavy and thus the momentum $\kv'$ and $\kv$ in \Eqref{HsoFT} can be treated as independent. 
In contrast, the momentum and the frequency carried by the spin gauge field and rotation matrix, $q$, $q'$, $\Omega$ and $\Omega'$, are small compared to $k$ and $k'$ considering an adiabatic regime.
The Hamiltonian of spin-orbit interaction then reads
\begin{align}
\Hso & \simeq  -i \sum_{\kv\kv'\qv} \sum_{ijk}  \epsilon_{ijk} \vso^{(k)}(\kv'-\kv)(k'-k)_i   \sum_{\omega\Omega}
\lt[ \frac{1}{2}(k'+k)_j R_{kl}(\qv,\Omega) a^\dagger_{\kv,\omega} \sigma_l a_{\kv',\omega+\Omega} 
  \rt. \nnr 
& \lt. 
    + \sum_{\qv'\Omega'} R_{kl}(\qv-\qv',\Omega-\Omega') \As{j}{l}(\qv',\Omega') 
           a^\dagger_{\kv,\omega} a_{\kv',\omega+\Omega}  
    \rt].
\end{align}

%
\begin{figure}\centering
\includegraphics[width=0.5\hsize]{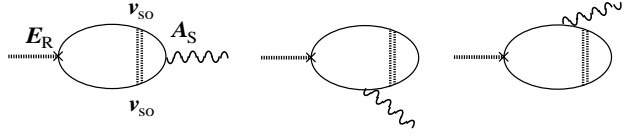}
\caption{
The Feynman diagrams describing the corrections of the Rashba-induced spin motive force  due to the spin relaxation,  $\Fv^{{\rm (sr)},A}$.
These contributions turn out to be higher order in the spin gauge field, and thus neglected compared with the contribution of Fig. \ref{FIGforce_sr}.
\label{FIGforce_srA}}
\end{figure}
%
%
Let us proceed to estimate the correction of the spin motive force due to the random spin-orbit interaction.
Since the bare spin motive force without relaxation is described by the Feynman diagram shown in Fig. \ref{FIGforce_A}, one would guess that the relaxation correction is represented by the processes shown in Fig. \ref{FIGforce_srA}.
These processes, however turn out to vanish at the first order in the gauge field as we will now demonstrate.

The sum of the three contributions in Fig. \ref{FIGforce_srA}, defined as $F^{{\rm (sr)},A}$,  
in the adiabatic limit ($\qv\ra0$)  is 
\begin{align}
F^{{\rm (sr)},A}_i &= -i\frac{2m\Deltasd}{\hbar \nel} \sum_{ijklmn} \sum_{\alpha\beta\gamma}
\epsilon_{ijk} \epsilon_{klm}
\alphaR_j n_l R_{mn} \sum_{\Omega_1,\Omega_2,\Omega_3}e^{i\Omega t}
 \As{0}{\gamma}(\Omega_3)
\sum_{\kv\kv'} 
\nnr
& \times
\sum_{abcdef} \epsilon_{abc}\epsilon_{def}k_b k_f k'_c k'_e 
R_{a\alpha}(\Omega_1)R_{d\beta}(\Omega_2)
\average{  \vso^{(a)}(\kv'-\kv) \vso^{(d)}(-\kv'+\kv) }
\nnr
& \times 
\sum_{\omega} 
\tr\lt[
 \sigma_n \gf{\kv\omega}{} 
 \sigma_\gamma \gf{\kv,\omega+\Omega_3}{} \sigma_\alpha  \gf{\kv',\omega+\Omega_1+\Omega_3}{} \sigma_\beta  \gf{\kv,\omega+\Omega}{} 
\rt. \nnr
& 
+
 \sigma_n \gf{\kv\omega}{} 
 \sigma_\alpha  \gf{\kv',\omega+\Omega_1}{} \sigma_\beta
  \gf{\kv,\omega+\Omega_1+\Omega_2}{}  \sigma_\gamma 
\gf{\kv,\omega+\Omega}{} 
\nnr
& \lt.
+ 
  \sigma_n \gf{\kv\omega}{} 
 \sigma_\alpha  \gf{\kv',\omega+\Omega_1}{} \sigma_\gamma  \gf{\kv',\omega+\Omega_1+\Omega_3}{} \sigma_\beta  \gf{\kv,\omega+\Omega}{} 
\rt]^<,\label{FsrA1}
\end{align}
where $\Omega\equiv \Omega_1+\Omega_2+\Omega_3$.
The average of the random potential is calculated as \cite{TE08}
\begin{align}
\average{  \vso^{(a)}(\kv'-\kv) \vso^{(d)}(-\kv'+\kv) } 
  = \average{ \vso^2 } \delta_{ad},
\label{Vsoaverage}
\end{align}
where $\average{ \vso^2 }$ represents the squared average of the spin-orbit strength.
Using the rotational symmetry, we replace $k_b k_f\ra\frac{1}{3}k^2 \delta_{bf}$ in \Eqref{FsrA1}.
The trace in the spin space is calculated using the following identities.
\begin{align}
\sum_\alpha \sigma_\alpha g \sigma_\alpha &= g+2\overline{g}
\nnr
\sum_\alpha \sigma_\alpha g \sigma_\gamma f \sigma_\alpha
   &= 2\delta_{\gamma z}\sigma_z(gf-\overline{g}\overline{f}) -g\sigma_\gamma f,
\end{align}
where
$g= g_0+\sigma_z g_1$ is any diagonal matrix, $\overline{g} \equiv  g_0-\sigma_z g_1$ and similarly for $f$. 
The result is 
\begin{align}
F^{{\rm (sr)},A}_i &= \frac{m\Deltasd}{\hbar \nel} 
\lt(\gamma_1^{{\rm (sr)},A} (\bm{\alphaR}\times\dot{\nv})
 +\gamma_3 ^{{\rm (sr)},A}(\bm{\alphaR}\times(\nv\times\dot{\nv})) \rt),
\end{align}
where
\begin{align}
\gamma_3^{{\rm (sr)},A} & = i\hbar\frac{2}{9}\average{\vso^2} \int\frac{d\omega}{2\pi} \sum_{\kv\kv'\pm} (\pm)
k^2(k')^2 \lt[ 
 f_{\omega+\Omega} 
   [ \gf{\kv\omega\pm}{\ret} \gf{\kv,\omega+\Omega,\mp}{\ret} (\gf{\kv',\omega+\Omega,\mp}{\ret} +2\gf{\kv',\omega+\Omega,\pm}{\ret} )  \gf{\kv,\omega+\Omega,\mp}{\ret} 
\rt. \nnr
&
+
\gf{\kv\omega\pm}{\ret} (\gf{\kv',\omega+\Omega,\pm}{\ret} +2\gf{\kv',\omega+\Omega,\mp}{\ret} )  \gf{\kv,\omega,\pm}{\ret} \gf{\kv,\omega+\Omega,\mp}{\ret}
-
\gf{\kv\omega\pm}{\ret} \gf{\kv',\omega,\pm}{\ret} \gf{\kv',\omega+\Omega,\mp}{\ret} \gf{\kv,\omega+\Omega,\mp}{\ret}  ]
+f_{\omega} [\ret\leftrightarrow\adv]  \nnr
&+ 
(f_{\omega+\Omega} -f_{\omega} )
[ \gf{\kv\omega\pm}{\ret} \gf{\kv,\omega+\Omega,\mp}{\adv} (\gf{\kv',\omega+\Omega,\mp}{\adv} +2\gf{\kv',\omega+\Omega,\pm}{\adv} )  \gf{\kv,\omega+\Omega,\mp}{\adv} 
\nnr
& \lt.
+
\gf{\kv\omega\pm}{\ret} (\gf{\kv',\omega+\Omega,\pm}{\ret} +2\gf{\kv',\omega+\Omega,\mp}{\ret} )  \gf{\kv,\omega,\pm}{\ret} \gf{\kv,\omega+\Omega,\mp}{\adv}
-
\gf{\kv\omega\pm}{\ret} \gf{\kv',\omega,\pm}{\ret} \gf{\kv',\omega+\Omega,\mp}{\adv} \gf{\kv,\omega+\Omega,\mp}{\adv}  ]
\rt]  ,
\end{align}
and $\gamma_1^{{\rm (sr)},A}$ is defined without the sign ($\pm$) in the summation.
The term $\gamma_1^{{\rm (sr)},A}$ gives only a small correction to \Eqref{Fbare}, and  we are here interested only in $\gamma_3^{{\rm (sr)},A}$.
In the adiabatic limit $\Omega\ra0$, the leading contribution turns out to be
\begin{align}
\gamma_3^{{\rm (sr)},A} &
\simeq  i\frac{4}{9} \int\frac{d\omega}{2\pi} \sum_{\kv\kv'\pm} (\pm)
(f_{\omega+\Omega} -f_{\omega} )
[2 \gf{\kv\omega\pm}{\ret} (\gf{\kv,\omega,\mp}{\adv})^2 \gf{\kv',\omega,\pm}{\adv}
 - \gf{\kv\omega\mp}{\ret} (\gf{\kv,\omega,\pm}{\adv})^2 \gf{\kv',\omega,\pm}{\adv}
]
\nnr
& =O(\Omega).
\end{align}
The contribution of $\gamma_3^{{\rm (sr)},A}$ to a force is therefore a second order in the time derivative as $\bm{\alphaR}\times\delo{t} (\nv\times\dot{\nv})$.
It is neglected in the adiabatic regime, and 
the processes shown in Fig. \ref{FIGforce_srA} do not thus to contribute to a perpendicular spin electric field proportional to  $\bm{\alphaR}\times (\nv\times\dot{\nv})$.

\begin{figure}\centering
\includegraphics[width=0.25\hsize]{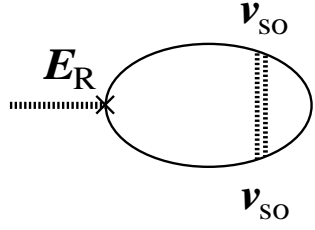}
\caption{
The Feynman diagrams describing the spin motive force arising from the Rashba interaction ($\EvR$) and spin relaxation, represented by the random spin-orbit interaction, $v_{\rm so}$.
Although  there is no explicit interaction with  spin gauge field ($\As{}{}$) in the diagram, the contribution turns out to results in a force containing the spin gauge field, proportional to 
 $\bm{\alphaR}\times (\nv\times\dot{\nv})$ \cite{Kohno07}.
\label{FIGforce_sr}}
\end{figure}

We will now show that dominant contribution arises from simpler processes shown in  Fig. \ref{FIGforce_sr}.
In fact, in those processes, the rotation matrices, $R_{ij}$, at the two vertices of the random spin-orbit interaction are defined in general at different space and time coordinates.
At equal time,  $\sum_{j} R_{ij}(t)R_{jk}(t)= \delta_{ik}$, and this relation results in a vanishing contribution to the force, while non-trivial contribution is expected at different time, since 
$\sum_{j} R_{ij}(t)R_{jk}(t')\neq \delta_{ik}$ if $t\neq t'$.
(In the frequency notation,  $R_{ij}$ carries a finite frequency, and thus
 $\sum_{j} R_{ij}(\Omega_1)R_{jk}(\Omega_2)\neq \delta_{ik}$ if $\Omega_1\neq\Omega_2$.) 
This fact in fact leads to a perpendicular contribution as we will now demonstrate.
The contribution of Fig. \ref{FIGforce_sr}, $F^{{\rm (sr)}}$, 
after averaging over random potential according to Eq. (\ref{Vsoaverage}) is 
\begin{align}
F^{{\rm (sr)}}_i &= -i\frac{4}{9}\frac{m\Deltasd}{\hbar \nel} \sum_{ijklmn} \sum_{\alpha\beta\gamma}
\epsilon_{ijk} \epsilon_{klm} \alphaR_j n_l R_{m\gamma} \sum_{\Omega_1,\Omega_2}e^{i\Omega t}
\sum_{\kv\kv'} 
\nnr
& \times
k^2(k')^2  R_{n\alpha}(\Omega_1)R_{n\beta}(\Omega_2) \average{  \vso^2 }
\nnr
& \times 
\sum_{\omega} 
\tr\lt[
 \sigma_\gamma \gf{\kv\omega}{} \sigma_\alpha \gf{\kv',\omega+\Omega_1}{} \sigma_\beta  \gf{\kv,\omega+\Omega}{} \rt]^<,  \label{FIIdef}
\end{align}
where $\Omega=\Omega_1+\Omega_2$.
It is easy to see that the contribution from the trace evaluated at $\Omega_1=\Omega_2=0$ vanishes, and the leading contribution arises from those at the linear order.
Summing the contribution obtained by replasing $\alpha$ with $\beta$ and $\Omega_1$ with $\Omega_2$, we can rewrite Eq. (\ref{FIIdef}) as 
\begin{align}
F^{{\rm (sr)}}_i &= -i\frac{4}{9}\frac{m\Deltasd}{\hbar \nel} \sum_{ijklmn} \sum_{\alpha\beta\gamma}
\epsilon_{ijk} \epsilon_{klm} \alphaR_j n_l R_{m\gamma} \sum_{\Omega_1,\Omega_2}e^{i\Omega t}
\sum_{\kv\kv'} 
\nnr
& \times
k^2(k')^2  R_{n\alpha}(\Omega_1)R_{n\beta}(\Omega_2) \average{  \vso^2 }
\nnr
& \times 
\sum_{\omega} 
\tr\lt[
 \sigma_\gamma \gf{\kv\omega-\frac{\Omega}{2}}{} \sigma_\alpha \gf{\kv',\omega+\Delta\Omega}{} \sigma_\beta  \gf{\kv,\omega+\frac{\Omega}{2}}{} 
+
 \sigma_\gamma \gf{\kv\omega-\frac{\Omega}{2}}{} \sigma_\beta \gf{\kv',\omega-\Delta\Omega}{} \sigma_\alpha  \gf{\kv,\omega+\frac{\Omega}{2}}{} 
\rt]^<,  \label{FII1}
\end{align}
where $\Delta\Omega=\frac{1}{2}(\Omega_1-\Omega_2)$.
 We note here that the contribution to \Eqref{FII1} linear in $\Omega(=\Omega_1+\Omega_2)$ vanishes because of $\sum_n R_{n\alpha}(t)R_{n\beta}(t)=\delta_{\alpha\beta}$.
We can thus expand \Eqref{FII1} with respect to $\Delta\Omega$ at $\Omega=0$.
We also know that the contribution asymmetric with respect to $\alpha$ and $\beta$ contributes to the force.
The result of the force is therefore
\begin{align}
F^{{\rm (sr)}}_i &= -i\frac{4}{9}\frac{m\Deltasd}{\hbar \nel} \sum_{ijklmn} \sum_{\alpha\beta\gamma}
\epsilon_{ijk} \epsilon_{klm} \alphaR_j n_l R_{m\gamma} \sum_{\Omega_1,\Omega_2}e^{i\Omega t} (\Omega_1-\Omega_2)
\sum_{\kv\kv'} 
\nnr
& \times
k^2(k')^2  R_{n\alpha}(\Omega_1)R_{n\beta}(\Omega_2) \average{  \vso^2 }
\hbar \int \frac{d\omega}{2\pi}  \nnr
& \times 
 \lt[
f'(\omega)\tr\lt[
 \sigma_\gamma \gf{\kv\omega}{\ret} 
 \sigma_\alpha (\gf{\kv',\omega}{\adv}-\gf{\kv',\omega}{\ret}) \sigma_\beta 
 \gf{\kv,\omega}{\adv}   - (\alpha\leftrightarrow\beta) \rt] \rt.
\nnr
& \lt.
+
\frac{f(\omega)}{2}
\tr\lt[ \lt[
 \sigma_\gamma \gf{\kv\omega}{\ret} 
 \sigma_\alpha (\gf{\kv',\omega}{\ret})^2 \sigma_\beta 
 \gf{\kv,\omega}{\ret} -(\ret\leftrightarrow\adv) \rt] 
-(\alpha\leftrightarrow\beta) \rt] \rt].
\label{FsrII3}
\end{align}
The summations over $\Omega_1$ and $\Omega_2$ are carried out as
\begin{align}
i\sum_{\Omega_1,\Omega_2} & e^{i\Omega t}  (\Omega_1-\Omega_2)
 R_{n\alpha}(\Omega_1)R_{n\beta}(\Omega_2) 
=\dot{R}_{n\alpha}(t)     R_{n\beta}(t)
    -{R}_{n\alpha}(t)\dot{R}_{n\beta}(t).
\label{sumOmOm}
\end{align}
Carrying out the summation over the index $n$, 
\begin{align}
\sum_n (\dot{R}_{n\alpha}(t)R_{n\beta}(t)-{R}_{n\alpha}(t)\dot{R}_{n\beta}(t))
&=
4(\dot{m}_\alpha m_\beta- m_\alpha\dot{m}_\beta).
\end{align}
Trace in \Eqref{FsrII3} gives rise to terms proportional to
$\epsilon_{\alpha\beta z}\delta_{\gamma z}$, $\epsilon_{\alpha\beta\gamma}$ and to
$\delta_{\alpha\gamma}\delta_{\beta z}- \delta_{\beta\gamma}\delta_{\alpha z}$.
The contribution containing $\delta_{\gamma z}$ vanishes, since it is proportional to 
$\sum_{lm}\epsilon_{klm}n_l R_{mz}=\sum_{lm}\epsilon_{klm}n_l n_{m}=0$.
The contribution containing $\epsilon_{\alpha\beta\gamma}$ leads to (by use of \Eqref{sumOmOm})
\begin{align}
i\sum_{\Omega_1,\Omega_2}e^{i\Omega t} (\Omega_1-\Omega_2)
& \sum_{lm}\epsilon_{klm}\epsilon_{\alpha\beta\gamma} 
     n_l R_{mz}(t) R_{n\alpha}(\Omega_1)R_{n\beta}(\Omega_2) 
\nnr
&=
-8(\dot{m}_k m_z + m_k \dot{m}_z) =-4\del{n_k}{t}.
\end{align}
Similarly,
the contribution containing 
$\delta_{\alpha\gamma}\delta_{\beta z}- \delta_{\beta\gamma}\delta_{\alpha z}$ leads to 
\begin{align}
i\sum_{\Omega_1,\Omega_2}e^{i\Omega t} (\Omega_1-\Omega_2)
 & \sum_{lm}\epsilon_{klm}(\delta_{\alpha\gamma}\delta_{\beta z}- \delta_{\beta\gamma}\delta_{\alpha z})
     n_l R_{mz}(t) R_{n\alpha}(\Omega_1)R_{n\beta}(\Omega_2) 
\nnr
&= -4\lt(\nv\times \del{\nv}{t}\rt)_k.
\end{align}
Defining 
\begin{align}
\frac{16}{9} \average{v_{\rm so}^2} \sum_{\kv\kv'} k^2(k')^2 
& \int \frac{d\omega}{2\pi}  
 \tr\lt[
f'(\omega)\lt[
 \sigma_\gamma \gf{\kv\omega}{\ret} 
 \sigma_\alpha (\gf{\kv',\omega}{\adv}-\gf{\kv',\omega}{\ret}) \sigma_\beta 
 \gf{\kv,\omega}{\adv}   - (\alpha\leftrightarrow\beta) \rt] \rt.
\nnr
& \lt.
+
\hbar\frac{f(\omega)}{2}
\lt[ \lt[
 \sigma_\gamma \gf{\kv\omega}{\ret} 
 \sigma_\alpha (\gf{\kv',\omega}{\ret})^2 \sigma_\beta 
 \gf{\kv,\omega}{\ret} -(\ret\leftrightarrow\adv) \rt] 
-(\alpha\leftrightarrow\beta) \rt] \rt]
\nnr
& \equiv
\gamma_1^{\rm (sr)} \epsilon_{\alpha\beta\gamma} +\gamma_2^{\rm (sr)} \epsilon_{\alpha\beta z}\delta_{\gamma z} 
+\gamma_3^{\rm (sr)} (\delta_{\alpha\gamma}\delta_{\beta z}- \delta_{\beta\gamma}\delta_{\alpha z})
,
\end{align}
the force is (the contribution from the term $\gamma_2^{\rm (sr)}$ vanishes) 
\begin{align}
F^{{\rm (sr)}}_i &= \frac{m\Deltasd}{\hbar \nel} 
(\gamma_1^{\rm (sr)} (\alphaRv\times \dot{\nv})+\gamma_3^{\rm (sr)}  (\alphaRv\times (\nv\times\dot{\nv}) ) ).
\end{align}
Here again we focus on the term $\gamma_3^{\rm (sr)}$, which reads
\begin{align}
\gamma_3^{\rm (sr)}
&= 
- \frac{16}{9\pi}\average{\vso^2} \sum_{\kv\kv'\pm} k^2(k')^2 
 (\gf{\kv',\pm}{\adv}-\gf{\kv',\pm}{\ret})
 (\gf{\kv\pm}{\ret}\gf{\kv,\mp}{\adv} - \gf{\kv\mp}{\ret}\gf{\kv,\pm}{\adv} ).
\end{align}
Summation over $\kv$, $\kv'$ are carried out as 
\begin{align}
\sum_{\kv'} (k')^2 
 (\gf{\kv',\pm}{\adv}-\gf{\kv',\pm}{\ret})
&= 
2\pi i \dos_\pm (k_{{\rm F}\pm})^2  \nnr
\sum_{\kv} k^2 
 \gf{\kv,\pm}{\ret} \gf{\kv,\mp}{\adv}
&= 
\frac{\pi i}{2\Deltasd}\sigma (\dos_+ (k_{{\rm F}+})^2 + \dos_- (k_{{\rm F}-})^2) ,
\end{align}
where $\dos_\pm$ and $k_{{\rm F}\pm}$ are the density of states and Fermi wave vector of the conduction electron with spin $\pm$, respectively.
The result is thus 
\begin{align}
\gamma_3^{\rm (sr)}
&= 
-\frac{32\pi}{9\Deltasd}\average{\vso^2} (\dos_+ (k_{{\rm F}+})^2 + \dos_- (k_{{\rm F}-})^2)^2 .
\end{align}
The same diagram as in Fig. \ref{FIGforce_sr} was discussed in the context of the Gilbert damping in  Ref. \cite{Kohno07} in the case of relaxation due to random magnetic impurities.
The result of $\gamma_3^{\rm (sr)}$  is in agreement with their result.

\section{Application \label{SEC:dw} }
In this section, we apply our results of spin motive force. 
We investigate three cases, a precessing uniform magnetization, domain wall motion, and current-induced magnetization switching, to explore the effect of spin electric field in detail.
Adding the contribution from the adiabatic gauge field \cite{Volovik87,Duine09}, 
the total spin electric field is written as 
\begin{align}
\Es{i} &=
\gamma \lt[ \nv\cdot(\dot{\nv}\times\nabla_i\nv)+ \beta_{\gamma} \dot{\nv}\cdot\nabla_i\nv \rt]
+
\gamma_{\rm R} \lt[ (\alphaRv \times\dot{\nv}) + \beta_{\rm R} (\alphaRv \times(\nv\times\dot{\nv}) ) \rt]_i   ,
\label{totalSEF}
\end{align}
where $\gamma$ and $\gamma_{\rm R}$  represent  coupling to the adiabatic gauge field and to the Rashba field, respectively. 
In the strong coupling limit, $\gamma=\frac{\hbar }{2e}$ and 
$\gamma_{\rm R}=\frac{m}{e\hbar}$.
Coefficients $\beta_{\gamma}$ and $\beta_{\rm R}$ represent the strength of the spin relaxation on spin electric field.
These constants are of the same order as the $\beta$ coefficient of the spin-transfer torque.

\subsection{Precession of uniform magnetization}

We first consider the spin electric field driven by a precessing uniform magnetization.
The effect of perpendicular component, ${\Evs}_{\perp}$, was studied in Ref. \cite{Takeuchi12}.
The adiabatic gauge field contribution, the term proportional to $\gamma$ in \Eqref{totalSEF}, vanishes identically for a uniform magnetization.
We consider a film or a thin wire with a perpendicular easy-axis anisotropy along $z$ axis.
The Rashba field is also along $z$ axis ($\alphaRv=\alphaR \hat{\zv}$) \cite{Miron10}. 
We consider the magnetization precessing with a constant tilt angle $\theta$. 
The direction is represented by a unit vector
\begin{align}
\nv=(\sin\theta\cos\phi(t),\sin\theta\sin\phi(t),\cos\theta).
\end{align}
Its time derivative is $\dot{\nv}=\sin\theta \dot{\phi}\ev_\phi$, and the result of spin electric field, ${\Evs}={\Evs}^{\rm R}$, is
\begin{align}
{\Evs}^{\rm R} &=
-\gamma_{\rm R} \alphaR \dot{\phi} \sin\theta \sqrt{1+(\beta_{\rm R})^2\cos^2\theta} 
   \lt(\begin{array}{c} \cos(\phi+\varphi) \\ \sin(\phi+\varphi) \\ 0 \end{array}  \rt)  ,
\end{align}
where $\varphi\equiv \sin^{-1}\frac{\beta_{\rm R}\cos\theta}{\sqrt{1+(\beta_{\rm R})^2\cos^2\theta}}$.
In the uniform magnetization case, contribution from spin relaxation (${\Evs}_{\perp}$ proportional to  $\beta_{\rm R}(\ll1)$) is small compared with ${\Evs}_{\parallel}$.
The magnitude of the electric field is thus estimated to be 
$|\Es{}^{\rm R}|=\gamma_{\rm R} \alphaR \omega \sin\theta$, where $\omega$ is the angular frequency of precession.
Considering a case of strong Rashba interaction, $\alphaR=3\times 10^{-10}$eV$\cdot$m \cite{Kim12}, the field is estimated for a frequency of $\nu=\omega/(2\pi)=100$MHz as 
$|\Es{}^{\rm R}|=2\times10^{3}$V/m.
For a film with size of 100nm, the voltage expected across the film is 0.2mV.
By detecting this AC spin motive force, therefore, one can verify the existence  of the Rashba interaction and estimate its strength in films. 
A great advantage of using spin motive force is that the voltage signal accumulates over the system as far as it is single domain, in contrast to the domain wall case, where the signal arises only from inside the domain wall.
Rashba-induced spin motive force would thus be very useful to distinguish the Rashba effect on the spin-transfer torque and spin Hall effect, which is an urgent issue at present \cite{Liu11,Liu12,WangRashba12}.

A DC component of voltage signal generated by a magnetization precession in a system of Pt film in contact with a permalloy film was observed in Ref. \cite{Saitoh06}, and the result was explained based on a spin pumping and inverse spin Hall effects.
The AC component of the voltage may contain Rashba-induced motive force if there is Rashba interaction at the permalloy-Pt interface.
Bilayer or multilayer systems with a perpendicular easy axis anisotropy like in Ref. \cite{Miron10} is suitable for detection of Rashba-induced spin motive force, since the effect of inverse spin Hall effect is neglected. 
This is because the spin polarization $\sigmav$ of spin current pumped  by the magnetization is parallel to the spin current flow $\jsv$ and thus inverse spin signal, proportional to $\jsv\times\sigmav$ \cite{Saitoh06}, vanishes.

\subsection{Moving domain wall}

Let us consider the effect of the spin electric field on moving Bloch domain wall in a thin film with a perpendicular easy axis anisotropy.
The film is in the $xy$ plane, and so the easy axis is along $z$ direction.
The Rashba field, $\EvR=E_{\rm R}\hat{\zv}$, is chosen along $z$ direction, as would be the case in experiments \cite{Miron10}.
The magnetization unit vector is written by use of polar coordinates as
\begin{align}
\nv=(\sin\theta\cos\phi,\sin\theta\sin\phi,\cos\theta).
\end{align}
A planar domain wall with magnetization changing along $x$ direction and at position $x=X(t)$ is described by a solution   
$\sin\theta=\frac{1}{\cosh\frac{x-X}{\lambda}}$ \cite{TKS_PR08},
where $\lambda$ is wall thickness.
treating $\phi$ as uniform inside the wall, we thus obtain
$\nabla_x{\nv}=-\frac{\sin\theta}{\lambda}\ev_\theta$ and 
\begin{align}
\dot{\nv}=\sin\theta\lt(\dot{\phi}\ev_{\phi}+\frac{\dot{X}}{\lambda}\ev_\theta\rt),
\end{align}
where $\ev_\phi=(-\sin\phi,\cos\phi,0)$ and $\ev_\theta=(\cos\theta\cos\phi,\cos\theta\sin\phi,-\sin\theta)$.
The total spin electric field averaged inside the domain wall 
(using $\average{\sin\theta}=\frac{2}{\pi}$ and $\average{ \cos\theta }=0$) reads
\begin{align}
{\Evs}   &=
\gamma \frac{1}{2\lambda}  \lt(\dot{\phi} -\beta \frac{\dot{X}}{\lambda} \rt) 
           \lt( \begin{array}{c} 1 \\ 0 \\ 0 \end{array} \right)
-\frac{2\gamma_{\rm R}}{\pi} E_{\rm R} \lt(\dot{\phi}+\beta_{\rm R}\frac{\dot{X}}{\lambda}\rt) 
 \lt( \begin{array}{c} \cos\phi \\ \sin\phi \\ 0 \end{array} \right) .
\label{EsDW}
\end{align}
The first contribution without Rashba interaction was discussed previously by Lucassen et al. \cite{Lucassen11}.
We see that the contribution proportional to wall speed, $\dot{X}$, arises solely from spin relaxation 
(contributions containing $\beta$ or $\beta_{\rm R}$), while $\dot{\phi}$ contribution arises without spin relaxation.
In the case of a steady spin-torque motion with oscillating $\phi$ \cite{TK04}, 
the Rashba contribution, proportional to $\cos\phi$ or $\sin\phi$ leads to AC component as predicted by Kim et al. \cite{Kim12}, while 
the adiabatic gauge field contribution ($\gamma$ contribution) has a DC component if $\dot{\phi}$ and $\dot{X}$ have finite average.
Experimental detection of Rashba-driven AC contribution rotating in the $xy$-plane provides a novel mean to confirm the dynamics of the wall's internal degrees of freedom, $\phi$, after the Walker's breakdown in the field-driven case and in the current-driven motion in the intrinsic pinning regime \cite{TK04,Koyama11}.

Let us study our result, \Eqref{EsDW}, in more detail considering the case without extrinsic pinning potential.
As described in Ref. \cite{TKS_PR08}, the equation of motion in this case is (neglecting oder of $\alpha^2,\alpha\beta \ll1$, where $\alpha$ is the Gilbert damping constant)
\begin{align}
\delp{X}{t} &= v_c\lt(P \jtil+\sin2\phi\rt) \nnr
\delp{\phi}{t} &= \frac{v_c}{\lambda}\lt((\beta-P\alpha)\jtil-\alpha \sin 2 \phi \rt),
\end{align}
where $P$ is spin polarization of the current, $v_c=\frac{\Kp\lambda S}{2\hbar}$ ($\Kp$ is hard axis anisotropy energy and $S$ is the magnitude of spin).

The Rashba contribution to the spin electric field is governed by a factor of (neglecting $O(\beta^2)$)
\begin{align}
  \lt(\dot{\phi} +\beta_{\rm R} \frac{\dot{X}}{\lambda} \rt) 
=\frac{v_c}{\lambda}\lt((\beta+P(\beta_{\rm R}-\alpha))\jtil+(\beta_{\rm R}-\alpha) \sin 2 \phi  \rt).
\end{align}
The Rashba contribution to the spin electric field is therefore
\begin{align}
{\Evs}^{\rm R}   &=
-\frac{2\gamma_{\rm R}}{\pi} E_{\rm R} \frac{v_c}{\lambda}
 \lt[
(\beta+P(\beta_{\rm R}-\alpha))\jtil
+2(\beta_{\rm R}-\alpha) \sin\phi \cos\phi \rt]
\lt( \begin{array}{c} \cos\phi \\ \sin\phi \\ 0 \end{array} \rt) .
\label{EsRDW}
\end{align}
By observing the AC component, parameters $(\beta_{\rm R}-\alpha) $ and $(\beta+P(\beta_{\rm R}-\alpha))$ are accessible experimentally.
Note that the spin relation contribution proportional to $\beta_{\rm R}$ (the fourth term of \Eqref{totalSEF})  contributes to the same order of magnitude as the bare contribution
(the third term of \Eqref{totalSEF}), since $\dot{X}/\lambda$ is larger than $\dot{\phi}$ by a factor of $\beta^{-1}$ \cite{TKS_PR08}.

\subsection{Threshold current for magnetization switching of a perpendicular layer}

For a single domain state, the total spin electric field (Eq. (\ref{totalSEF})) is reduced to  

\begin{equation}
{\bm E}_s^\pm = \pm \frac{m \alpha_R}{e \hbar} [(\bm{\hat{z}} \times \bm{\dot n})+\beta_R (\bm{\hat{z}} \times (\bm{n} \times \bm{\dot n}))]. 
\end{equation}
This spin electric field generates a spin current density via 
${\bm j}_s = \sigma_\uparrow {\bm E}_s^+ - \sigma_\downarrow {\bm E}_s^- = \sigma_c {\bm E}_s^+$ where $\sigma_{\uparrow (\downarrow)}$ is the electrical conductivity of majority (minority) electrons and $ \sigma_c = \sigma_\uparrow + \sigma_\downarrow$. The spin transfer torque due to ${\bm j}_s$ can be obtained by inserting ${\bm j}_s$ into Rashba adiabatic and nonadiabatic spin torques \cite{Obata08,Manchon09,WangRashba12,KimRashba12,Pesin12} (neglecting order of $\beta_R \beta$) 

\begin{equation}
{\bm T}({\bm j}_s) =  \gamma_g \sigma_c \left(\frac{m \alpha_R}{e \hbar} \right)^2 [\bm{n} \times (\tilde {\it{D}} \cdot \bm{\dot n})+\beta_R \bm{n} \times (\bm{\hat{z}} \times (\tilde {\it{D}} \cdot \bm{\dot n}))-\beta \bm{n} \times \bm{n} \times (\tilde {\it{D}} \cdot \bm{\dot n})], \label{T}
\end{equation}
where $\gamma_g $ is the gyromagnetic ratio, and $\tilde {\it{D}}$ is a 3 by 3 tensor with the element $\it{D}_{ij} = \delta_{ij}(1-\delta_{iz})$. Then magnetization dynamics with no current injection is described by the modified Landau-Lifshitz-Gilbert equation,

\begin{equation}
{\bm{\dot n}}=-\gamma_g {\bm n} \times {\bm H}_{eff}+\alpha {\bm n} \times {\bm{\dot n}} + \tilde{\alpha} \bm{n} \times (\tilde {\it{D}} \cdot \bm{\dot n})+\tilde{\alpha}[\beta_R \bm{n} \times (\bm{\hat{z}} \times (\tilde {\it{D}} \cdot \bm{\dot n}))-\beta \bm{n} \times \bm{n} \times (\tilde {\it{D}} \cdot \bm{\dot n})] , \label{LLG}
\end{equation}
where ${\bm H}_{eff}$ is the effective magnetic field acting on ${\bm n}$
and $\tilde{\alpha}= \gamma_g \sigma_c (m \alpha_R/e \hbar)^2$. One finds that the third term on R.H.S. of Eq. (\ref{LLG}) has the same form with the Gilbert damping torque and thus makes the damping anisotropic \cite{Kim12}.

We next discuss the role of the last term on the R.H.S. of Eq. (\ref{LLG}), which is related to $\beta_R$ and $\beta$. For a perpendicularly magnetized layer where magnetization undergoes a small-angle precession around the easy axis, $\bm n$ and ${\bm H}_{eff}$ are described by $(m_0 \cos (\omega t), m_0 \sin (\omega t), 1)$ and $(-N_{xy}M_s m_0 \cos (\omega t), -N_{xy}M_s m_0 \sin (\omega t), H_k - N_z M_s)$, respectivley, where $m_0\ll1$, $\omega=2\pi f$, $f$ is the precession frequency, and $N_{xy} (N_z)$ is the in-plane (out-of-plane) demagnetization factor. For this case, the precession frequency $f$ is given by $f = \frac{\gamma_g}{2\pi} \frac{H_k-(N_z-N_{xy})M_S}{1-\tilde {\alpha}(\beta-\beta_R)}$ and thus the last term on the R.H.S. of Eq. (\ref{LLG}) may be interpreted as a modification of the gyromagnetic ratio induced by spin electric field.

Finally we investigate effect of Rashba spin-orbit coupling on current-induced magnetization switching in perpendicular spin valves, which is of considerable interest for ultrahigh density spin transfer torque magnetic random access memories (STT-MRAMs). To find a threshold current for perpendicular STT-MRAM of which free layer is subject to Rashba spin-orbit coupling, one uses Eq. (\ref{LLG}) with including the Slonczewski's spin torque term, $\gamma_g \frac{\hbar \eta j_e }{2 e M_S t_F} {\bm n} \times ( {\bm n} \times \bm{\hat{z}})$, where $\eta$ is the spin torque efficiency factor, $j_e$ is the charge current density, $M_S$ is the saturation magnetization, and $t_F$ is the thickness of free layer. Assuming a small-angle precession (i.e., $m_0\ll1$) and using the energy dissipation rate $\left< \partial E/\partial t \right> $ (i.e., $\left< \partial E/\partial t \right> = -M_s \left< {\bm H}_{eff} \cdot \partial {\bm n}/\partial t \right>=0$, where $\left<...\right>$ is a time-average over a period), one finds a threshold current density $j_{SW}$ as

\begin{equation}
j_{SW} = (\alpha + \tilde{\alpha}) \frac{2e}{\hbar} \frac{M_S t_F}{\eta}(H_k - (N_z-N_{xy})M_S). \label{Jsw}
\end{equation}
From Eq. (\ref{Jsw}), one finds that Rashba spin-orbit coupling enhances the damping by $\tilde{\alpha}$ and consequently increases $j_{SW}$. On the other hand, the modified gyromagnetic ratio does not affect $j_{SW}$ because  $j_{SW}$ is entirely determined by the competition between the damping torque and the Slonczewski's spin torque. One may want to reduce $j_{SW}$ as much as possible for application of perpendicular STT-MRAM. Eq. (\ref{Jsw}) suggests that Rashba spin-orbit coupling should be minimized for this purpose.

\section{Summary}
We have calculated the spin motive force acting on the electron induced by the  Rashba interaction and sd interaction 
 based on a microscopic formulation. 
We have shown that there are contributions in the limit of strong sd interaction; one proportional to $\EvR\times\dot{\nv}(\equiv {\Evs}_{\parallel}^{\rm R})$, which arises without spin relation, and the other perpendicular component proportional to 
$\EvR\times(\nv\times\dot{\nv})(\equiv {\Evs}_{\perp}^{\rm R})$ arising from the spin relaxation
($\EvR$ and $\nv$ are the Rashba electric field and localized spin direction, respectively).

Rashba-induced field arises even in the uniform magnetization case, in contrast to the spin electric field arising from adiabatic gauge field, which is finite only if magnetization  has both spatial structure (finite $\nabla\nv$) and  time-dependence. 
Measurement of spin motive force driven by uniform magnetization precession would be useful in confirming the existence of the Rashba interaction in multilayer systems with perpendicular easy axis like the system studied by Miron \cite{Miron10}.

In view of our result, the appearance of a perpendicular spin motive force in the weak sd coupling limit  discussed in the context of spin damping monopole \cite{Takeuchi12} is naturally understood, since spin relaxation is strong in the weak coupling regime having no particular spin quantization axis.
There is a possibility that the other spin electric field ${\Evs}_{\parallel}^{\rm R}$ also contains monopoles. 
This possibility is currently under study by investigating the spin magnetic field in the strong coupling limit.

\section*{Acknowledgment}
The authors thank H. Kohno for valuable discussions. 
This work was supported by 
 a Grant-in-Aid for Scientific Research (B) (Grant No. 22340104) from Japan Society for the Promotion of Science and UK-Japanese Collaboration on Current-Driven Domain Wall Dynamics from JST. K.J.L acknowledges a support from the NRF (2010-0023798).


\bibliography{/home/tatara/References/12,/home/tatara/References/gt}

\end{document}